\documentclass{article}
\usepackage{spconf,amsmath,graphicx}
\usepackage{multirow}

\def\SP#1{\textsuperscript{#1}}
\def\SB#1{\textsubscript{#1}}

\title{Spectrogram-Channels U-Net: A source separation model\\viewing each channel as the spectrogram of each source}
\name{Jaehoon Oh$^*$, Duyeon Kim$^*$, Se-Young Yun
}
\address{$^*$Both authors contributed equally to this work.\\
Graduate School of Knowledge Service Engineering, KAIST\\
\{jhoon.oh,duyeon,yunseyoung\}@kaist.ac.kr
}
\begin{document}
%
\maketitle
\begin{abstract}
Sound source separation has attracted attention from Music Information Retrieval(MIR) researchers, since it is related to many MIR tasks such as automatic lyric transcription, singer identification, and voice conversion. In this paper, we propose an intuitive spectrogram-based model for source separation by adapting U-Net. We call it \textit{Spectrogram-Channels U-Net}, which means each channel of the output corresponds to the spectrogram of separated source itself. The proposed model can be used for not only singing voice separation but also multi-instrument separation by changing only the number of output channels. In addition, we propose a loss function that balances volumes between different sources. Finally, we yield performance that is state-of-the-art on both separation tasks.
\end{abstract}
\begin{keywords}
Source Separation, Singing Voice Separation, Multi-instrument Separation, Spectrogram-Channels U-Net
\end{keywords}
\section{Introduction}
\label{sec:intro}

\begin{figure*}[t]
\begin{center}
\includegraphics[width=17cm]{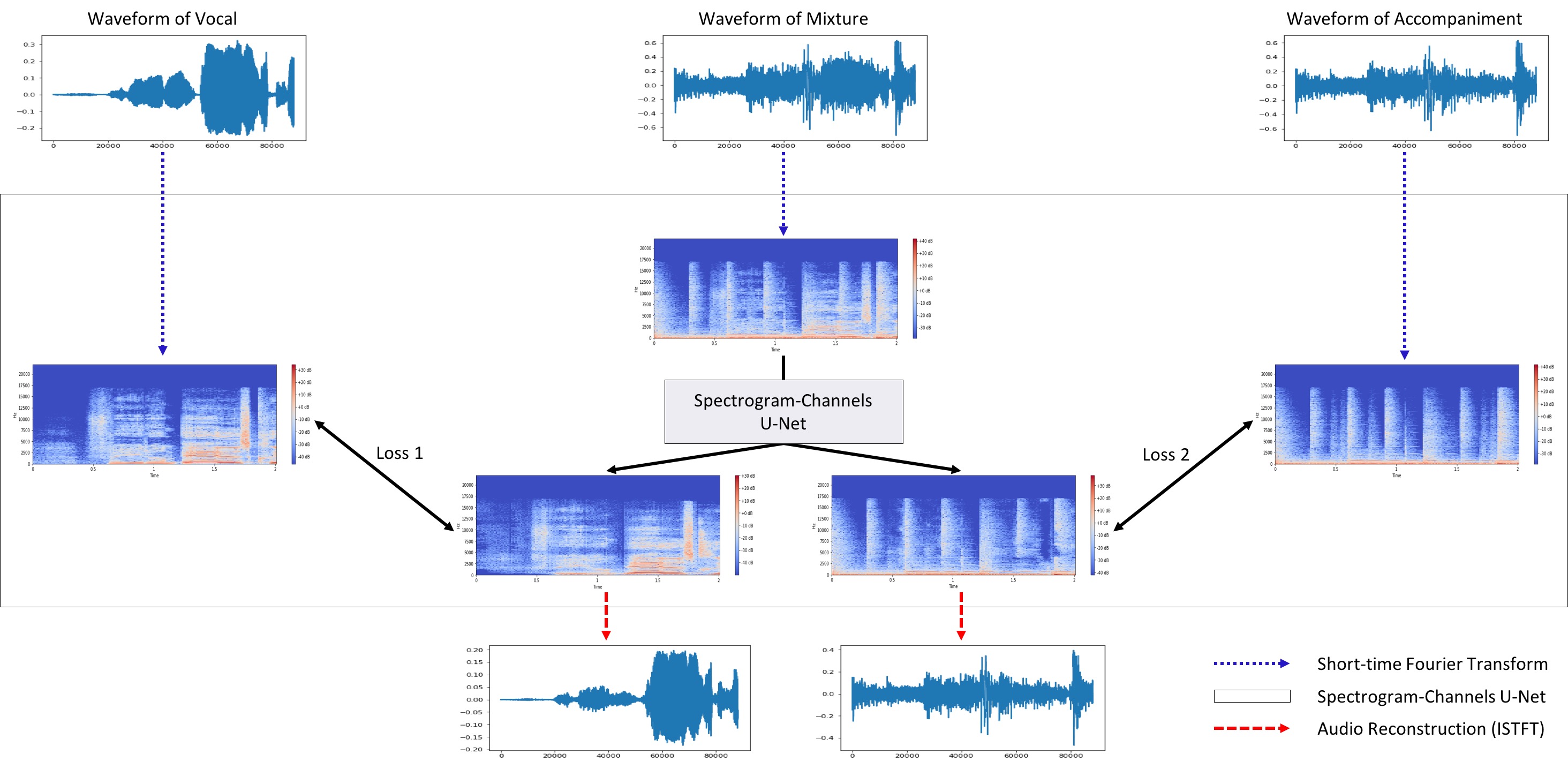}
\caption{Overall architecture for singing voice separation}
\label{fig:arch}
\end{center}
\end{figure*}
Separating sources from a song is an important basic task in Music Information Retrieval(MIR) because many tasks in MIR require a clean source. However, this is a difficult task because a song consists of many sources. If one can separate vocal and its accompaniment clearly, it opens up a new horizon in terms of music applications; automatic lyric transcription, singer identification, voice conversion, music similarity, genre classification, and so on.

In many previous studies, non-negative matrix factorization (NMF) was used for the source separation of the audio signal \cite{virtanen2007monaural, liang2013beta, souviraa2015music}. Bayesian modeling \cite{ozerov2007adaptation} and Principle Component Analysis (PCA) were also used \cite{huang2012singing}.

In recent years, deep learning techniques have shown outstanding performance in various fields, so many studies have tried to adopt deep learning techniques for audio signal source separation. Grais et al. \cite{grais2014deep} used both NMF and Deep Neural Networks(DNN) and showed better separation performance than just using NMF. Especially, Convolutional Neural Networks(CNN) showed fine performance in solving image-based problems \cite{krizhevsky2012imagenet}. CNN can extract features well by using convolutional filters and is efficient in aspect of memory because of weight sharing. Andreas Jansson et al. \cite{jansson2017singing} showed that U-Net can yield good performance on sound domain by using a spectrogram as an input. U-Net \cite{ronneberger2015u}, which is composed of a fully convolutional encoder and decoder network and features skip connection from encoder to decoder, was originally introduced in the biomedical field.

Besides CNN, there are other deep learning techniques studied in source separation. Huang et al. \cite{huang2014singing} used Recurrent Neural Networks(RNN) for singing voice separation. Takahashi et al. \cite{takahashi2018mmdenselstm} used both CNN and RNN for audio source separation. Another approach is using Generative Adversarial Networks (GAN) \cite{goodfellow2014generative} which is composed of a generator network that makes fake input and a discriminator network that determines whether the input is fake or real. Santiago et al. \cite{pascual2017segan} made Speech Enhancement GAN(SEGAN), which is the end-to-end GAN model applied to a domain where models filter out voice from noise environment. And Singing Voice Separation GAN(SVSGAN) \cite{fan2017svsgan} is the GAN model generating voice from music.

In this paper, we propose an intuitive 2D convolution neural network using spectrograms for sound source separation, \textit{Spectrogram-Channels U-Net}, which is based on the model in \cite{jansson2017singing} and uses 2-channels output for singing voice separation or 4-channels output for multi-instrument separation. This is the main difference from the original U-Net model for sound separation. The original model has a drawback in that the output is a mask, which is multiplied to a mixture spectrogram and can generate only a separated vocal spectrogram. To separate multi-instrument sources at the same time, additional models are needed. Our output remedies this since it is not a mask but a direct spectrogram, which enables us to separate multiple sources simultaneously. Secondly, we propose a weighted loss function to balance volumes between different sources since every source would have quite different volumes from each other. We finally evaluate our model using an objective measure and show reasonable performance on both tasks.

\label{sec:proposed}
\section{Proposed Model}

\begin{figure}[h!]
\begin{center}
\includegraphics[width=0.7\columnwidth]{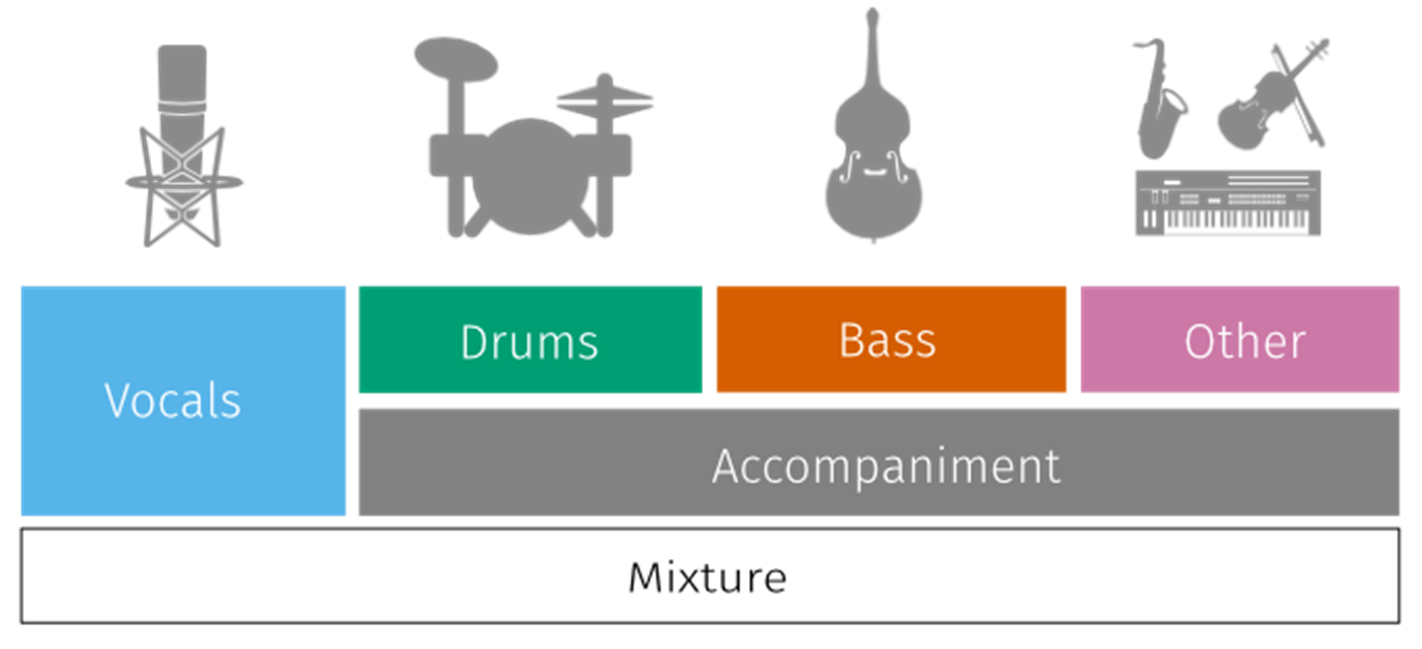}
\caption{The components of the dataset}
\label{fig:dt}
\end{center}
\end{figure}

We use the \textit{musdb18} dataset, which is a representative dataset for sound source separation. This includes isolated vocals and accompaniment. Accompaniment is also provided as decomposed drums, bass, and other sources. Figure \ref{fig:dt} shows components of dataset \cite{musdb18}. There are 100 songs for training and 50 songs for testing. We do not do any additional processing other than converting stereo to mono. Input data is mixture for both separation tasks. For singing voice separation, target data are vocals and accompaniment. While for multi-instrument separation, target data are vocals, drums, bass, and other.

Figure \ref{fig:arch} represents our overall architecture for singing voice separation. It is composed of three parts: Short-time Fourier Transform, Spectrogram-Channels U-Net, and Audio Reconstruction.

\subsection{Short-time Fourier Transform}\label{Feature Extraction}
We transform a waveform into a spectrogram to make input and target data of the network by the following procedure:
\begin{description}
\item[$\bullet$ Step 1:]We sample waveform of songs at 44,100Hz. Then, we cut songs into 2 second increments.
\item[$\bullet$ Step 2:]We apply a short-time Fourier transform (STFT) to waveform signals with a window size of 2048 and hop length of 512 frames. As a result, we get the complex-valued spectrograms of signals.
\item[$\bullet$ Step 3:]We decompose spectrograms into magnitude and phase. The former is used as input and targets to the network, the latter is used for audio reconstruction. For the sake of simplicity, spectrogram means magnitude in this paper.
\end{description}

\subsection{Spectrogram-Channels U-Net}\label{sec:DCEDNs}
\begin{figure}[h!]
\begin{center}
\includegraphics[width=0.9\columnwidth]{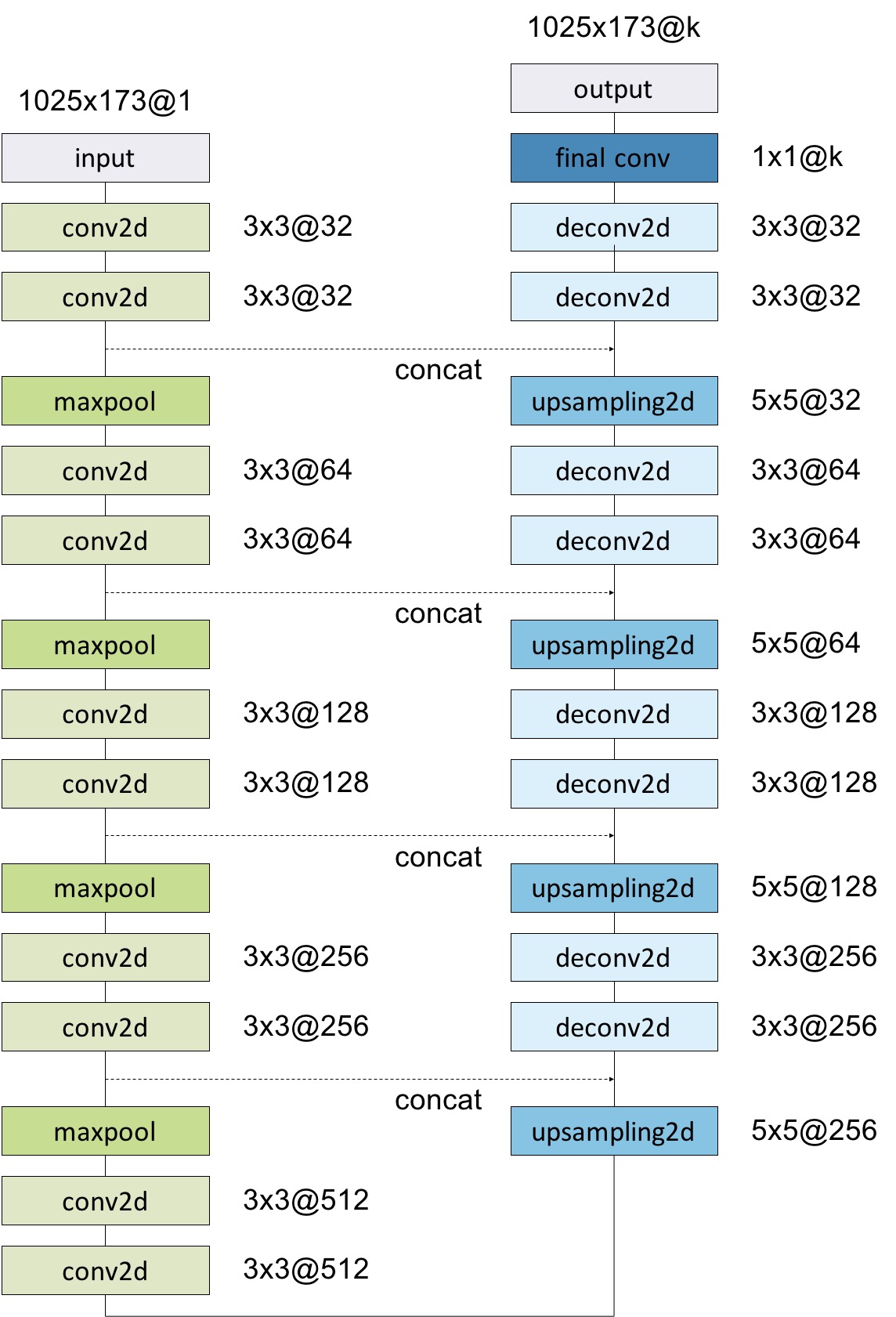}
\caption{The detail of \textit{Spectrogram-Channels U-Net}}
\label{fig:str}
\end{center}
\end{figure}

Figure \ref{fig:str} represents the detail of \textit{Spectrogram-Channels U-Net}. The network consists of the encoder part and the decoder part. The encoder part has a repeated structure of two components: a convolution layer that keeps spectrogram size and increases the number of channels, and a max pooling layer that decreases spectrogram size in half and keeps the number of channels. Each convolution layer in the encoder consists of convolution of kernel size 3x3 and padding 1, batch normalization, and rectified linear units(ReLU) activation function.

The decoder part also has a repeated structure of two components: an upsampling deconvolution layer that doubles spectrogram size and decreases the number of channels in half, and a deconvolution layer that keeps spectrogram size and decreases the number of channels. The former consists of deconvolution of kernel size 5x5, stride 2, and padding 2, batch normalization, ReLU activation function, and dropout with probability 0.4. The latter consists of deconvolution of kernel size 3x3 and padding 1, batch normalization, and ReLU activation function.

The encoder and decoder, having the same spectrogram size, are concatenated via skip connection. Since skip connection delivers the information of mixture from encoder to decoder, the combination of channels on the decoder part is also related to mixture.

From this idea, we make the difference from \cite{jansson2017singing} in which they used the concept of mask for separation. Rather, we make multiple channels output and view each channel of this output as the spectrogram of each source directly. For this idea, we use 1x1 convolution as the final convolution and use ReLU activation function since the values of spectrograms are non-negative.

In summary, the input of this model is spectrogram of mixture, and each channel of the output corresponds to the spectrogram of each source that we want to separate. For convenience, let $source$ be the spectrogram of a target and $channel\SB{source}$ be the spectrogram of a separated source from a mixture. For example, in $L(vocal, channel\SB{vocal})$, $vocal$ is the spectrogram of vocal and $channel\SB{vocal}$ is the spectrogram of separated vocal from a mixture. For the training, we define loss functions as follow for each task:
\begin{center}
$loss$(Singing Voice) = 
$\alpha*L(vocal, channel\SB{vocal})$\\
\hspace{43pt}$+(1-\alpha)*L(acc, channel\SB{acc})$
\end{center}
\begin{center}
$loss$(Multi-instrument) = 
$\alpha\SB{1}*L(vocal, channel\SB{vocal})$\\
\hspace{97pt}$+\alpha\SB{2}*L(drum, channel\SB{drum})$\\
\hspace{88pt}$+\alpha\SB{3}*L(bass, channel\SB{bass})$\\
\hspace{95pt}$+\alpha\SB{4}*L(other, channel\SB{other})$\\
$(\alpha\SB{1}+\alpha\SB{2}+\alpha\SB{3}+\alpha\SB{4}=1)$
\end{center}
where $L$ is the average of L1 losses on every pixel. We train the network using ADAM optimizer with weight decay 1e-6, and batch size of 8. We perform 20 epochs with learning rate of 1e-3, then perform 20 epochs more with learning rate of 1e-4.

\subsection{Audio Reconstruction}
We re-transform spectrogram into waveform to listen to the reconstructed sound and to evaluate. We multiply the phase of mixture and the magnitude of separated source elementwisely where mixture is the input and separated source is the output. After that, we get the waveform using inverse STFT.

\section{Experiments}
\label{sec:experiments}
To measure source separation performance objectively \cite{vincent2006performance}, we use Signal-to-Distortion Ratio(SDR) of separated sources. The formula is as follows:
\begin{center}
SDR $:= 10\log\SB{10}\frac{||s\SB{target}||\SP{2}}{||e\SB{interf}+e\SB{noise}+e\SB{artif}||\SP{2}}$ \\
\end{center}
where $s\SB{target}$ is a version of modified by an allowed distortion for some signal $s\SB{j}$ and $e\SB{interf}$, $e\SB{noise}$, and $e\SB{artif}$ are, respectively, the interferences, noise, and artifacts error terms. To reduce artifacts by STFT and ISTFT, we use 20 seconds as an input of model and then evaluate it every one second.

\subsection{Singing Voice Separation}
For singing voice separation, we make 4 different models according to $\alpha$ in the loss function. \textbf{M1} sets $\alpha$ as 1.0, which means we extract only vocal source using this model. Then, we get accompaniment source by submitting extracted vocal source from mixture in waveform. On the other hand, \textbf{M2} uses 0.0, which means we extract only accompaniment source using this model. Then, we get vocal source as the same way in \textbf{M1}. \textbf{M3} uses 0.5, which means we extract vocal source from first output channel and accompaniment source from second output channel. Finally, \textbf{M4} uses 0.707, which is the weight balancing between vocal and accompaniment. This value is calculated in waveform. First, we calculate the mean of 2-norm of vocal training set and accompaniment training set respectively. We get the $\alpha$ value using this formula; $\alpha*$(the mean of 2-norm of vocal) = $(1-\alpha)*$(the mean of 2-norm of accompaniment). The reason why the weight of vocal source is greater is because vocal has smaller volume of sound than accompaniment. The balancing weight will help the sources that have small volume to be separated better than before.

The results of singing voice separation are shown in Table \ref{table:svs}. Because each of \textbf{M1} and \textbf{M2} aims to extract vocal and accompaniment source respectively, it can be seen that \textbf{M1} shows best score in vocal separation, and \textbf{M2} shows almost best score in accompaniment separation compared to others. And we can see that the score of vocal source separation of \textbf{M4} has increased compared to \textbf{M3}. This is because the balancing weight enhances the vocal source separation.

\begin{table}[]
\begin{center}
\begin{tabular}{|c|c|c|c|c|c||c|}
\hline
                       &        &M1   &M2   &M3   &M4   &C1\\ \hline
                       & Med.   &\textbf{5.93} &0.68 &5.79 &5.90 &4.46  \\ \cline{2-7} 
                       & MAD    &6.11 &6.96 &6.14 &6.16 &3.21  \\ \cline{2-7} 
                       & Mean   &\textbf{2.50} &-2.32&2.28 &2.45 &0.65  \\ \cline{2-7} 
\multirow{-4}{*}{Voc.} & SD     &11.09&9.90 &11.00&11.09&13.67 \\ \hline
                       & Med.   &11.43&11.82&\textbf{11.93}&11.67&10.69 \\ \cline{2-7} 
                       & MAD    &5.01 &5.49 &5.18 &5.13 &3.15  \\ \cline{2-7} 
                       & Mean   &\textbf{13.47}&\textbf{13.47}&13.36&12.85&11.85 \\ \cline{2-7} 
\multirow{-4}{*}{Acc.} & SD     &8.49 &7.37 &7.09 &6.66 &7.03  \\ \hline
\end{tabular}
\caption{The results of singing voice separation}
\label{table:svs}
\end{center}
\end{table}

\subsection{Multi-instrument Separation}
For multi-instrument separation, we make 2 different models according to $\alpha's$ in the loss function. \textbf{M5} uses [0.25, 0.25, 0.25, 0.25], which means we extract each source using the same weights. On the other hand, \textbf{M6} uses [0.297, 0.262, 0.232, 0.209], which means we extract each source using balancing weights. The way to get balancing weights is the same as singing voice separation case; First, we calculate the ratio between $\alpha's$ using $\alpha\SB{1}*$(the mean of 2-norm of vocal) = $\alpha\SB{2}*$(the mean of 2-norm of drum) = $\alpha\SB{3}*$(the mean of 2-norm of bass) = $\alpha\SB{4}*$(the mean of 2-norm of other), then get exact values using $\alpha\SB{1}+\alpha\SB{2}+\alpha\SB{3}+\alpha\SB{4}=1$. From this balancing weights, we can infer conversely that other is the biggest source and vocal is the smallest source. 

The results of multi-instrument separation are shown in Table \ref{table:mis}. Since \textbf{M5} and \textbf{M6} separate each four sources, the scores of vocal source separation are lower than singing voice separation. In result of \textbf{M6} using balancing weights, we can see that the separation scores of vocal, drum, and even bass, which have relatively lower sound volume than other, have increased compared to \textbf{M5}. It can be seen that the balancing weights adjust the ratio between several sources to some extent.

\begin{table}
\begin{center}
\begin{tabular}{ |c|c|c|c||c| }
\hline
                       &        &M5    &M6    &C2   \\ \hline
                       & Med.   &5.32  &\textbf{5.40}  &3.00 \\ \cline{2-5} 
                       & MAD    &5.95  &5.89  &2.76 \\ \cline{2-5} 
                       & Mean   &1.99  &\textbf{2.11}  &-2.10\\ \cline{2-5} 
\multirow{-4}{*}{Vocal}& SD     &10.49 &10.40 &15.41\\ \hline
                       & Med.   &4.84  &\textbf{5.05}  &4.15 \\ \cline{2-5} 
                       & MAD    &4.60  &4.46  &1.99 \\ \cline{2-5} 
                       & Mean   &4.13  &\textbf{4.34}  &2.88 \\ \cline{2-5} 
\multirow{-4}{*}{Drum} & SD     &6.59  &6.31  &7.68 \\ \hline
                       & Med.   &2.44  &2.66  &\textbf{2.91} \\ \cline{2-5} 
                       & MAD    &5.55  &5.53  &2.47 \\ \cline{2-5} 
                       & Mean   &0.96  &\textbf{1.14}  &-0.30\\ \cline{2-5} 
\multirow{-4}{*}{Bass} & SD     &7.89  &7.88  &13.50\\ \hline
                       & Med.   &\textbf{2.70}  &2.54  &2.03 \\ \cline{2-5} 
                       & MAD    &4.00  &4.00  &1.64 \\ \cline{2-5} 
                       & Mean   &\textbf{2.34}  &2.06  &1.68 \\ \cline{2-5} 
\multirow{-4}{*}{Other}& SD     &6.00  &5.85  &6.14 \\ \hline
\end{tabular}
\caption{The results of multi-instrument separation}
\label{table:mis}
\end{center}
\end{table}

\subsection{Comparison with Others Using MUSDB18}
We compare ours with others reported in \cite{stoter20182018}. A various methods are used for enhancing performance; additional 800 songs for training \cite{takahashi2018mmdenselstm, uhlich2017improving}, data augmentation techniques such as random scaling and mixing with \textit{musdb18} \cite{takahashi2018mmdenselstm, uhlich2017improving, liu2018denoising, stoller2018wave}, and Wiener filter as a post-processing \cite{takahashi2018mmdenselstm, uhlich2017improving, liu2018denoising}. We make models without any method mentioned above. We bring the results C1 in Table \ref{table:svs} and C2 in Table \ref{table:mis} from \cite{stoller2018wave}. Since they used random scaling only for data augmentation, it is the most suitable for comparison. These results show that our model outperforms this model except a bass source on a multi-instrument separation task.

\section{Conclusion}
\label{sec:conclusion}
We propose an intuitive \textit{Spectrogram-Channels U-Net} viewing each channel of output as the spectrogram of each source directly. The proposed model can be adapted for both separation tasks by only changing the number of output channels. In addition, to consider the balance between sources, we use an weighted loss function by calculating the means of 2-norm on each source training set. Finally, we evaluate our model and earn comparable results on both tasks without data augmentation and post-processing. For future work, we will investigate other balancing ways besides the proposed balancing method, or 2-norm mean. By adapting more appropriate a balancing method, we could get better sound source separation.

\newpage
\bibliographystyle{IEEEbib}
\bibliography{refs}

\begin{thebibliography}{10}

\bibitem{virtanen2007monaural}
Tuomas Virtanen,
\newblock ``Monaural sound source separation by nonnegative matrix
  factorization with temporal continuity and sparseness criteria,''
\newblock {\em IEEE transactions on audio, speech, and language processing},
  vol. 15, no. 3, pp. 1066--1074, 2007.

\bibitem{liang2013beta}
Dawen Liang, Matthew~D Hoffman, and Daniel~PW Ellis,
\newblock ``Beta process sparse nonnegative matrix factorization for music.,''
\newblock in {\em ISMIR}, 2013, pp. 375--380.

\bibitem{souviraa2015music}
Nathan Souvira{\`a}-Labastie, Emmanuel Vincent, and Fr{\'e}d{\'e}ric Bimbot,
\newblock ``Music separation guided by cover tracks: Designing the joint nmf
  model,''
\newblock in {\em Acoustics, Speech and Signal Processing (ICASSP), 2015 IEEE
  International Conference on}. IEEE, 2015, pp. 484--488.

\bibitem{ozerov2007adaptation}
Alexey Ozerov, Pierrick Philippe, Frdric Bimbot, and Rmi Gribonval,
\newblock ``Adaptation of bayesian models for single-channel source separation
  and its application to voice/music separation in popular songs,''
\newblock {\em IEEE Transactions on Audio, Speech, and Language Processing},
  vol. 15, no. 5, pp. 1564--1578, 2007.

\bibitem{huang2012singing}
Po-Sen Huang, Scott~Deeann Chen, Paris Smaragdis, and Mark Hasegawa-Johnson,
\newblock ``Singing-voice separation from monaural recordings using robust
  principal component analysis,''
\newblock in {\em Acoustics, Speech and Signal Processing (ICASSP), 2012 IEEE
  International Conference on}. IEEE, 2012, pp. 57--60.

\bibitem{grais2014deep}
Emad~M Grais, Mehmet~Umut Sen, and Hakan Erdogan,
\newblock ``Deep neural networks for single channel source separation,''
\newblock in {\em Acoustics, Speech and Signal Processing (ICASSP), 2014 IEEE
  International Conference on}. IEEE, 2014, pp. 3734--3738.

\bibitem{krizhevsky2012imagenet}
Alex Krizhevsky, Ilya Sutskever, and Geoffrey~E Hinton,
\newblock ``Imagenet classification with deep convolutional neural networks,''
\newblock in {\em Advances in neural information processing systems}, 2012, pp.
  1097--1105.

\bibitem{jansson2017singing}
Andreas Jansson, Eric Humphrey, Nicola Montecchio, Rachel Bittner, Aparna
  Kumar, and Tillman Weyde,
\newblock ``Singing voice separation with deep u-net convolutional networks,''
\newblock 2017.

\bibitem{ronneberger2015u}
Olaf Ronneberger, Philipp Fischer, and Thomas Brox,
\newblock ``U-net: Convolutional networks for biomedical image segmentation,''
\newblock in {\em International Conference on Medical image computing and
  computer-assisted intervention}. Springer, 2015, pp. 234--241.

\bibitem{huang2014singing}
Po-Sen Huang, Minje Kim, Mark Hasegawa-Johnson, and Paris Smaragdis,
\newblock ``Singing-voice separation from monaural recordings using deep
  recurrent neural networks.,''
\newblock in {\em ISMIR}, 2014, pp. 477--482.

\bibitem{takahashi2018mmdenselstm}
Naoya Takahashi, Nabarun Goswami, and Yuki Mitsufuji,
\newblock ``Mmdenselstm: An efficient combination of convolutional and
  recurrent neural networks for audio source separation,''
\newblock {\em arXiv preprint arXiv:1805.02410}, 2018.

\bibitem{goodfellow2014generative}
Ian Goodfellow, Jean Pouget-Abadie, Mehdi Mirza, Bing Xu, David Warde-Farley,
  Sherjil Ozair, Aaron Courville, and Yoshua Bengio,
\newblock ``Generative adversarial nets,''
\newblock in {\em Advances in neural information processing systems}, 2014, pp.
  2672--2680.

\bibitem{pascual2017segan}
Santiago Pascual, Antonio Bonafonte, and Joan Serra,
\newblock ``Segan: Speech enhancement generative adversarial network,''
\newblock {\em arXiv preprint arXiv:1703.09452}, 2017.

\bibitem{fan2017svsgan}
Zhe-Cheng Fan, Yen-Lin Lai, and Jyh-Shing~Roger Jang,
\newblock ``Svsgan: Singing voice separation via generative adversarial
  network,''
\newblock {\em arXiv preprint arXiv:1710.11428}, 2017.

\bibitem{musdb18}
Zafar Rafii, Antoine Liutkus, Fabian-Robert Stöter, Stylianos~Ioannis
  Mimilakis, and Rachel Bittner,
\newblock ``The {MUSDB18} corpus for music separation,'' Dec. 2017.

\bibitem{vincent2006performance}
Emmanuel Vincent, R{\'e}mi Gribonval, and C{\'e}dric F{\'e}votte,
\newblock ``Performance measurement in blind audio source separation,''
\newblock {\em IEEE transactions on audio, speech, and language processing},
  vol. 14, no. 4, pp. 1462--1469, 2006.

\bibitem{stoter20182018}
Fabian-Robert St{\"o}ter, Antoine Liutkus, and Nobutaka Ito,
\newblock ``The 2018 signal separation evaluation campaign,''
\newblock in {\em International Conference on Latent Variable Analysis and
  Signal Separation}. Springer, 2018, pp. 293--305.

\bibitem{uhlich2017improving}
Stefan Uhlich, Marcello Porcu, Franck Giron, Michael Enenkl, Thomas Kemp, Naoya
  Takahashi, and Yuki Mitsufuji,
\newblock ``Improving music source separation based on deep neural networks
  through data augmentation and network blending,''
\newblock in {\em Acoustics, Speech and Signal Processing (ICASSP), 2017 IEEE
  International Conference on}. IEEE, 2017, pp. 261--265.

\bibitem{liu2018denoising}
Jen-Yu Liu and Yi-Hsuan Yang,
\newblock ``Denoising auto-encoder with recurrent skip connections and residual
  regression for music source separation,''
\newblock {\em arXiv preprint arXiv:1807.01898}, 2018.

\bibitem{stoller2018wave}
Daniel Stoller, Sebastian Ewert, and Simon Dixon,
\newblock ``Wave-u-net: A multi-scale neural network for end-to-end audio
  source separation,''
\newblock {\em arXiv preprint arXiv:1806.03185}, 2018.

\end{thebibliography}

\end{document}